\documentclass[%
 reprint,
showpacs,preprintnumbers,
 amsmath,amssymb,
 aps,
]{revtex4-1}

\usepackage{graphicx}
\usepackage{dcolumn}
\usepackage{bm}

\newcommand{\Bx}     {\mbox{{\boldmath $x$}}}

\newcommand{\Bf}     {\mbox{{\boldmath $f$}}}
\newcommand{\Bs}     {\mbox{{\boldmath $s$}}}
\newcommand{\Br}     {\mbox{{\boldmath $r$}}}

\begin{document}


\title{Additional Bending of Light in  Sun's Vicinity \\ by its Interior  Index of Refraction}

\author{Jacob T. Fokkema}          \email{j.t.fokkema@tudelft.nl}

\author{Peter M. van den Berg}   \email{p.m.vandenberg@tudelft.nl}
\affiliation{%
Faculty of Applied Physics,
Delft University of Technology,
the Netherlands}%

\date{\today}
\begin{abstract}

In the seventies, scientists observed discrepancies of the bending of light around the Sun based on Einstein's prediction of the curvature of star light due to the mass of the Sun.  We claim that the interior electromagnetic properties of the Sun influence the curvature of the light path outside the Sun as well.
In this paper,
 we investigate the additional deflection of light in the vacuum region surrounding the Sun by its electromagnetic parameters.
   Starting with Maxwell's equations, we show how the deflection of light passing  the Sun  depends on the electric permittivity and the magnetic permeability of the interior of the Sun.
   The electromagnetic field equations in Cartesian coordinates are transformed to  ones in a Riemannian geometry defined through an appropriate metric.  The metric  is generated by the introduction of a refractional potential.  The geodetic lines  are constructed from this potential.
    As far as the deflection of light  propagating along these geodetic lines is concerned, we show that the existence of a refractional potential influences the light path outside any object with a typical refractive index.
    Our results add new aspects to the bending of star light explained by general relativity.
    Some astrophysical observations, which cannot be explained by gravity in a satisfactory manner, are justified by the electromagnetic model. In particular, the frequency dependency of the light deflection is discussed. We show that the additional bending due to the refractive index is proportional to the third power of the inverse distance. The general relativity predicts that the bending due to the mass is proportional to the inverse distance.

\end{abstract}

\maketitle
%
\section{Introduction}

Albert Einstein \cite{Einstein1911}  predicted the bending of light from a distant star by the mass of our Sun through the heaviness of light. The experimental verification in 1919  by Eddington, see \cite{Dyson1920}, of the apparent position shift of the star on the firmament, corroborated the theory of general relativity of Einstein \cite{Einstein1916}.
An overview of the 1919 measurements is given by  Will \cite{Will2015}.
 Einstein explained the bending of the grazing starlight by the gravity of the Sun as it followed the curved geodetic path of light in four-dimensional space. A total deflection angle of 1.75 arcsec is arrived at. Woodward and Yourgrau \cite{WoodwardYourgrau1970a,WoodwardYourgrau1970b}
discussed a paradox in the interaction of the gravitational and electromagnetic fields. To solve this paradox they introduced a frequency dependency of the speed of light in the gravitational field, while
Treder \cite{Treder1971} used a nonlinear generalizations of Maxwell's dynamics in the general relativity.
Merat {\em et al} \cite{Merat1974} explained the effect on the light deflection close to the vicinity of the solar limb by introducing a dispersive layer.

The leading question is: are Maxwell's equations able to explain the change of  the light path in the vicinity a refractive object?
This investigation is the aim of the present paper.
We consider Maxwell's equations with a Riemannian metric.  In this geometry we have a bounded object of general form and composition. Let us denote the fastest path of light waves as the geodetic line. We choose a non-trivial metric and we arrive at a tensor form of  the electromagnetic equations. We determine the geodetic lines  from the Helmholtz decomposition theorem for the spatial coordinate changes as a function of refractive index. This  leads to the introduction of a global refractional potential, which indeed influences the curvature of the geodetic line outside the object.

In this paper, the deflection of star light passing the Sun is discussed. We  consider  a radially inhomogeneous sphere model with a certain refractive index depending on the radial position. We derive a  relation, in which the total deflection angle is related to the mean value of the refractive index of the Sun. This refractive index is frequency dependent.

\section{Maxwell's equations in tensor notation}

Light is an electromagnetic phenomenon. We consider waves with complex time factor $\exp(-{\rm i}\omega t)$,
where ${\rm i}$ is imaginary unit, $\omega$ is the radial frequency and $t$ is the time.
 In a vacuum domain,  with Cartesian coordinates $\Bx \in{\cal R}^3$, we write Maxwell's equations in the frequency domain
as
\begin{equation}
\begin{array}{rcccl}
e_{ijk} \partial_j B_k &+& \displaystyle\frac{1}{c_0^2}{\rm i}\omega E_i &=&   0\,,
\\[4mm]
 e_{ijk}   \partial_j E_k & -&                      {\rm i}\omega B_i &=&  0\,,
\end{array}
\label{Maxwellvacuum}
\end{equation}
where  $E_j=E_j(\Bx,\omega)$ is the electric field vector, $B_j=B_j(\Bx,\omega)$ is the magnetic field vector,
$c_0$ is the velocity of light in vacuum, and $e_{ijk}$ is  the Levi-Civita symbol.
For repeated subscripts, Einstein's summation convention is used.

In a subdomain ${\cal S}$ of ${\cal R}^3$, containing a material medium, we define the  spatially dependent refractive index $n = n(\Bx,\omega) =c_0/c(\Bx,\omega)$.  Further $\mu=\mu(\Bx,\omega)$ represents the spatially dependent magnetic permeability. Note that the wave velocity is given by $c =1/\sqrt{\varepsilon\mu}$, where $\varepsilon =\varepsilon(\Bx,\omega)$ is the spatially dependent electric permittivity.  We neglect absorption, so that all material parameters are real valued.

Maxwell's equations in  ${\cal S}$ are given by
\begin{equation}
\begin{array}{rcccl}\displaystyle
\,e_{ijk}\, \mu\,\partial_j\Bigl(\frac{1}{\mu} B_k\Bigr) &+& \displaystyle\frac{n^2}{c_0^2}{\rm i}\omega E_i &=&   0\,,
\\[4mm]
  e_{ijk}   \partial_j E_k  &- &                    {\rm i}\omega B_i &=&  0\,.
\end{array}
 \label{Maxwellmatter}
\end{equation}

  For vacuum in the whole ${\cal R}^3$ we have $n=1$ and the constant  magnetic permeability $\mu=\mu_0$.   Then, the geodetic lines are straight.  In a vacuum domain outside ${\cal S}$ with a material medium, we are not allowed to conclude that the geodetic lines in that  domain are straight.
 The geodetic lines are not equivalent to the ray paths in optics, which are defined using  a high-frequency approximation of Maxwell's equations. In the neighborhood of domain ${\cal S}$ with $n\neq1$, these optical rays in vacuum remain  straight when they pass ${\cal S}$, because within the ray approximation the interaction with  matter in ${\cal S}$ is neglected. However, the presence of the object ${\cal S}$ leads to diffraction of the incident wave and this may influence the path of propagation. In fact, the geodetic line may become curved.
 Although, with the help of present-day computer codes a more or less complete solution of Maxwell's equations is possible, the structure of the geodetic lines is hardly to observe from the numerical solution. We therefore investigate the nature of Maxwell's  equations in a different coordinate system.

We introduce a Riemannian geometry with the symmetric metric tensor $g_{ij}$.
 In tensor  notation, Maxwell's equations are
\begin{equation}
\begin{array}{rcccl}\displaystyle
\,  g_{il}\,\epsilon^{ljk}\,\mu\, \overline{\partial}_j\Bigl(\frac{1}{\mu} \overline{B}_k\Bigr)& + &\displaystyle \frac{n}{c_0^2}\, \, {\rm i}\omega \overline{E}_i &=&   0\,,
 \\[4mm]\displaystyle
   g_{il}\,\epsilon^{ljk}  \overline{\partial}_j \Bigl(\frac{1}{n}\,\overline{E}_k\Bigr)  &-  &        {\rm i}\omega \overline{B}_i &=&  0\,,
   \end{array}
 \label{MaxwellStretched}
\end{equation}
 where ${\overline E}_i$,  ${\overline H}_i $  and $\overline{\partial}_j$ are the electric field vector, the magnetic field vector and the partial derivative  in the Riemannian geometry, respectively. These vectors are defined as
\begin{equation}
\Bigl\{\overline{B}_i,\overline{E}_i,\overline{\partial}_i\Bigr\}  =  \frac{\partial x^j}{\partial \overline{x}^i}
\Bigl\{B_j,n E_j,\partial_j\Bigr\} \,.
\end{equation}
The permutation  tensor $\epsilon^{ijk}$ is related to the Levi-Civita symbol as
\begin{equation}
\epsilon^{ijk} = \frac{1}{\sqrt{g}} \, e_{ijk}\,,
\label{permutation}
\end{equation}
where $g$ is the determinant of the metric tensor $g_{ij}$, given by
\begin{equation}
g_{ij} = \frac{\partial x^l}{\partial\overline{x}^i} \, \frac{\partial x^l}{\partial\overline{x}^j\,}.
\label{DefGeometricTensor}
\end{equation}

Using this definition in Eq.~(\ref{MaxwellStretched}), we obtain
\begin{equation}
\begin{array}{rcccl}\displaystyle
 \frac{g_{il}}{n\sqrt{g}} \,e_{ljk}\, \mu\,\overline{\partial}_j\Bigl(\frac{1}{\mu} \overline{B}_k\Bigr) &+&\displaystyle \frac{1}{c_0^2}\, \, {\rm i}\omega \overline{E}_i&=&   0\,, \\[4mm] \displaystyle
   \frac{g_{il}}{n\sqrt{g}}\,e_{ljk}\,n\,  \overline{\partial}_j \Bigl(\frac{1}{n}\overline{E}_k \Bigr) &   -  &          {\rm i}\omega  \overline{B}_i&=&  0\,.
   \end{array}
 \label{Maxwellmod}
\end{equation}
In a vacuum domain outside the Sun, these equations simplify to
\begin{equation}
\begin{array}{rcccl}\displaystyle
 \frac{g_{il}}{\sqrt{g}} \,e_{ljk}\overline{\partial}_j \overline{B}_k &+&\displaystyle \frac{1}{c_0^2}\, \, {\rm i}\omega \overline{E}_i&=&   0\,, \\[4mm] \displaystyle
   \frac{g_{il}}{\sqrt{g}}\,e_{ljk}  \overline{\partial}_j \overline{E}_k  &   -  &          {\rm i}\omega  \overline{B}_i&=&  0\,.
   \end{array}
 \label{MaxwellmodSimple}
\end{equation}
From the properties of the metric tensor being continuous and differentiable, we conclude
that the  tensor $g_{ij}= g_{ij}(n)$ is a global function of the refractive index $n= n(\Bx)$, for any point $\Bx$ in space.
Hence, the curvature in a vacuum domain outside the Sun is completely determined by the refractive  index of the interior of the Sun. A wave approaching the Sun is 'feeling'
the presence of the Sun before it has reached the Sun. In this way the Sun shows its emergence to the incident wave field, see also Feynman \cite{Feynman1964}, {\em A more precise statement of Fermat's principle}.

In the remainder of the paper, we omit the symbol $\omega$ to denote the frequency dependency of the field and material quantities. If necessary we only give their spatial dependency.

\section{SPECIFICATION OF THE GEOMETRIC TENSOR}

\subsection{Orthogonal transformation}

If we choose the simple orthogonal transformation
\begin{equation} \frac{\partial x^j}{\partial \overline{x}^i} =\frac{1}{n} \delta_i^j\,,
\quad \frac{\partial \overline {x}^i} {\partial x^j}= n\, \delta_j^i \,,
\label{g1-definition}
\end{equation}
where $\delta_{i}^j$ is the Kronecker tensor,  and the corresponding metric tensor is given by
\begin{equation}
\quad   g_{ij} = \frac{1}{n^2} \,\delta_{ij}\,.
\label{g-definition}
\end{equation}
It is obvious that this transformation is a local operator, viz., in a vacuum domain outside the Sun, it leads to propagation along a straight line in the Cartesian space.
Inside an inhomogeneous Sun, this transformation leads to curved paths according to the standard ray theory.

\subsection{Symmetric transformation}

To find the connection of this  local transformation to  a global extension, we choose
a symmetric transformation matrix (curl-free). The divergence of $\overline{x}^j(\Bx)$ (trace of the operator)
directly follows from the contraction of the second relation of Eq.~(\ref{g1-definition}) as
\begin{equation}
\frac{\partial \overline {x}^j} {\partial x^j}= 3n\,.
\label{divergence_f}
\end{equation}

We maintain these properties in the extension to a global transformation. To this end, we
define
 the position vectors in the Cartesian and Riemannian geometries as  $x^j$ and $\overline{x}^j$, respectively.
The  values of these coordinates in the Cartesian coordinate system are related to each other as
\begin{equation}
\overline{x}^j(\Bx) = x^j + f^j(\Bx)\,,
\label{geodetic1}
\end{equation}
in which $f^j$ is a continuous and differentiable vector field.
To obtain the new expression for the transformation tensor, we differentiate Eq.~(\ref{geodetic1}) with respect to the Cartesian coordinates and arrive at
\begin{equation}
\frac{\partial \overline {x}^j} {\partial x^i} = \delta^j_i + \frac{\partial  {f}^j} {\partial x^i}\,.
\label{NewTransform}
\end{equation}
At this point, we switch to the matrix representation of the tensors and introduce the curvature matrix ${\cal C}_{ij}$ as a representation of the transformation tensor, viz.,
\begin{equation}
{\cal C}_{ij}  = \delta_{ij} + \partial_i f_j\,.
\label{DEfCurv}
\end{equation}
To complete the specification of the extended transformation tensor,
 we contract Eq.~(\ref{NewTransform}) and use
Eq.~(\ref{divergence_f}). We then have
\begin{equation}
\partial_j f_j= 3(n-1)\,.
\label{geodetic3}
\end{equation}
Then, Helmholtz decomposition theorem for a curl-free field provides the non-trivial solution
\begin{equation}
f_j = - \,\partial_j\Phi\,,
\label{geodetic4}
\end{equation}
where we define $\Phi$ as the refractional potential, given by
\begin{equation}
\Phi(\Bx) = \int_{\Bx'\in {\cal S}} \frac{3\,[n(\Bx')-1]}{4\pi |\Bx-\Bx'|}\,{\rm d}V\,.
\label{PHI}
\end{equation}
Obviously, $f_j$ is the tension due to the difference in refractive index with respect to vacuum. We denote $f_j$  as the refractional tension.
This representation is valid under the condition that $n-1$ vanishes at the boundary surface of the domain
${\cal  S}$.

Before, we continue with our analysis, we conclude that Eqs.~(\ref{geodetic1}), (\ref{geodetic4}) and (\ref{PHI}) define our spatial transformation from $\Bx$ to $\overline{\Bx}$ coordinates.
This definition holds for any distribution of the refractive index  inside  domain ${\cal S}$.
Note that the expression of the refractive potential yields a non-zero value outside ${\cal S}$ and this confirms that
the refractive index distribution  inside the object ${\cal S}$ not only determines the spatial coordinate transformation inside this object, but also outside. Hence, the geodetic lines in the vacuum domain around ${\cal S}$ are influenced by the inner refractive index of the object.
It is obvious that $\overline{\Bx}$ is a nonlinear function of $\Bx$ and therefore it is difficult to find directly the geodetic lines.

Next we consider the scalar arclength $\overline{{\rm d}s}$ given by
\begin{equation}
\overline{{\rm d}s}= \left[\overline{{\rm d}x}_i\, \overline{{\rm d}x}_i\right]^{\frac{1}{2}}\,.
\label{geodetic11}
\end{equation}
Introducing the unit vector $\hat{s}_k= {\rm d}x_k / {\rm d}s$,  $\hat{s}_k\hat{s}_k =1$, we write $ \overline{{\rm d}s}= \overline{{\rm d}s}(\Bx,\hat{\Bs})$ as
 \begin{equation}
\overline{{\rm d}s}  = \left[{\cal C}_{jl}\,{\cal C}_{jk}\, \hat{s}_l\,\hat{s}_k\right]^\frac{1}{2} \,
{\rm d}s\, .
\label{geodetic15}
\end{equation}

To investigate the dynamic behavior, see p.~114  of Born and Wolf \cite{BornWolf1959}, we consider the optical length of the geodetic path, which is defined by the actual length of the path times the index of refraction.  Hence, the left-hand side of Eq.~(\ref{geodetic15}) represents the optical length of the path.
Therefore, we introduce
 the virtual refraction index ${n^{\rm g}}$ along the geodetic path as
\begin{equation}
 {n^{\rm g}}(\Bx,\hat{\Bs})=   \left[{\cal C}_{jl}\,{\cal C}_{jk}\, \hat{s}_l\hat{s}_k \right]^{\frac{1}{2}}\,.
\label{geodetic18}
\end{equation}
We remark that the actual computation of this refractive index is simplified by employing an eigenvalue decomposition of the matrix  $C_{jk}$. Using Eqs.~(\ref{DEfCurv}) and (\ref{geodetic3}) we observe that
the trace of this matrix  is equal to $3n$ .
 The trace of a linear operator is  the divergence of that linear operator with respect to its linear argument.
Since the matrix $C_{jk}$ is real and symmetric, an eigenvalue decomposition with positive eigenvalues exists and the sum of the eigenvalues is equal to the trace.
The procedure to determine the eigenvalues are discussed in our spherical example.

Basically, the virtual refractive index ${n^{\rm g}}(\Bx,\hat{\Bs})$  controls the path of the geodetic line in a similar way as the refractive index $n(\Bx)$  controls the path of optical rays. Note that the virtual refractive index is not  only determined by the local position of the geodetic line, but it also depends on the direction of the geodetic line at this position. We construct this geodetic line by considering the explicit Eulor integration  of the classic differential equation for the evolution of  an optical ray path, see p.~121 of Born and Wolf \cite{BornWolf1959}, but we replace the physical refractive index $n$ by the virtual counterpart ${n^{\rm g}}$, viz.
\begin{equation}
\frac{d[{n^{\rm g}}(\Bx,\hat{\Bs})\, \hat{s}_j]}{d{s}} = \partial_j\, {n^{\rm g}}\,, \quad {\rm with}\; \hat{s}_j =\frac{dx_j}{ds}\,,
\label{difeqray}
\end{equation}
where $x_j=x_j(s)$ is the trajectory of the geodetic line and $s$ is the parametric distance along this trajectory,
while $\hat{s}_j$ is the tangential unit vector along the geodetic line.
We note that this differential equation holds for refractive indices, which are invariant for the direction of the geodetic path. However, the explicit Euler integration  updates the ray position  and ray direction in such a way that only the previous information of position and direction is used over the pertaining path segment. During each integration step, the path directions are taken to be constant.

For a rotationally symmetric configuration the  present analysis simplifies.  For this specific case, we shall discuss  the construction of  the geodetic lines in full detail.

\section{Radially inhomogeneous medium }

In order to determine the eigenvalues of the curvature tensor for the spherical example,
we remark that the trace of $C_{jk}$  is equal to $3n$, where we used Eq.~(\ref{geodetic3}) The eigenvectors are spanned by the unit vectors in the directions of the tension $\Bf$. One of the eigenvalues corresponds to the eigenvector $f_j/|\Bf|$, so that this eigenvalue $\lambda$ satisfies
\begin{equation}
{\cal C}_{jk} \frac{f_k}{|\Bf|}= \lambda \frac{\,f_j}{|\Bf|}\,.
\end{equation}
Use of Eq.~(\ref{DEfCurv}) and contraction  of the result with $f_j/|\Bf|$  leads to
\begin{equation}
\lambda = 1 + \frac{f_k}{|\Bf|} \partial_k |\Bf|\,.
\label{eigenvaluetau}
\end{equation}

At this point we use spherical coordinates.
Then, the refractional potential and tension are determined in closed form (see Appendix A).
In this case, the  tension  depends on $R$ only and is directed in the radial direction.
Hence, $f_\theta = f_\phi =0$ and
the radial component is given by, see Eq.~(\ref{APPNfR}),
\begin{equation}
f_R(R)=  \frac{3}{R^2} \int_{0}^{R}[n(r)-1]\, r^2 \, {\rm d}r\,.
\label{sphere2}
\end{equation}
The Cartesian components of the tension are obtained as
\begin{equation}
f_k = \frac{{x_k}}{R} f_{R}(R)\,.
\label{sphere2cartesian}
\end{equation}

From Eq.~(\ref{eigenvaluetau}) it follows that the eigenvalue in the radial direction is given by
\begin{equation}
\lambda_R = 1+ \partial_R f_R\,,
\label{eigenVR}
\end{equation}
while the eigenvalues in the tangential directions follow from the trace
\begin{equation}
\lambda_R + \lambda_\theta + \lambda_\phi = 3n\,.
\end{equation}
 In view of the axial symmetry of our configuration, the tangential eigenvalues are the same. We therefore confine our analysis to the plane in which the geodetic path is defined.  Hence, we suffice with the computation of $\lambda_\theta$, viz.,
 \begin{equation}
\lambda_\theta  = (3n - \lambda_R) /2\,.
\end{equation}
 The eigenvalues depend only on $ \partial_R f_R$. From Eq.~(\ref{geodetic3}) we observe that
\begin{equation}
 \frac{1}{R^2}\partial_R (R^2f_R) = 3(n\!-\!1)\,,
\label{sphere4}
\end{equation}
or
\begin{equation}
  \partial_R f_R = 3(n-1)-\frac{2\,f_R}{R}\,.
\label{eigenVRmod}
\end{equation}
From Eq.~(\ref{eigenVR}) it follows that
\begin{equation}
\lambda_R = 1-2\frac{f_R}{R} + 3(n\!-\!1)
\end{equation}
and
\begin{equation}
\lambda_\theta = 1+ \frac{f_R}{R} \,.
\end{equation}

The virtual refractive index is then obtained as, cf. Eq.~(\ref{geodetic18}),
\begin{equation}
{n^{\rm g}} =\Bigl[\Bigl(1-2 \frac{f_R}{R}+ 3(n\!-\!1)\Bigr)^2 \hat{s}_R^2
+
\Bigl(1 + \frac{f_R}{R}\Bigr)^2 \hat{s}_\theta^2 \Bigr]^{\frac{1}{2}}\!,
\label{VIRTUALSIMP2}
\end{equation}
where $ \hat{s}_R=
\cos(\theta-\alpha)$ and $\hat{s}_\theta=\sin(\theta-\alpha)$ are the unit eigenvectors. Here, $\theta$ is the angle between  $\hat{\Br}$  and the $x_1$-direction, while $\alpha$ is the angle between $\hat{\Bs}$  and the $x_1$-direction.

\section{Geodetic lines outside an inhomogeneous sphere in vacuum}
We now consider a radially inhomogeneous sphere with radius $R_{\cal S}$. Let us define the mean value of the refractive index of the sphere as $n_{\cal S}$.
Outside the sphere, the  second expression  for the tension of Eq.~(\ref{sphere2}) simplifies to
 \begin{equation}
f_R(R)=(n_{\cal S}-1) \frac{R_{\cal S}^3}{R^2}\,, \quad  R>R_{\cal S}\,.
\label{tensionhomsphereGreater}
\end{equation}
We have computed some geodetic lines by solving the differential equation of Eq.~(\ref{difeqray}), using the method described below this equation. Using our deflection angle $\alpha$, this set of differential equations  in the plane $x_3=0$  is written as
\begin{equation}
\begin{array}{rcl}
\frac{\displaystyle d({n^{\rm g}}\, \sin\alpha)}{\displaystyle d{s}} \!\!&\! =\!&\!\!  \partial_2\, {n^{\rm g}}\,,\\ [2mm]
\frac{\displaystyle d({n^{\rm g}}\, \cos\alpha)}{\displaystyle d{s}} \!\!&\! =\!&\!\!  \partial_1\, {n^{\rm g}}\,.
\end{array}
\label{raydifeq}
\end{equation}
These equations holds for refractive indices, which do not depend on the direction of the geodetic path. To include the directional dependence on the direction of the path,
we solve this set of first order ordinary differential equations using Euler's method with step size $\Delta s$. At each step of this explicit scheme, the path direction  is determined by
the one of the previous step and it does not change over the path segment $\Delta s$.
In our linear approximation,
the rotational factors $\hat{s}_R^2$ and  $ \hat{s}_\theta^2$ do not change over each path segment and depend only on $R$. This means that
the spatial derivatives of the virtual refractive index in the plane $x_3=0$  are given by
\begin{equation}
\begin{array}{rcl}
\partial_1{n^{\rm g}}\!\!&\!= \!&\!\!\displaystyle \frac{x_1}{R} \partial_R n^{\rm g} =A\, \cos\theta\, /\,{n^{\rm g}}\,, \\[4mm]
\partial_2{n^{\rm g}}\!\!&\!=\!&\!\!\displaystyle \frac{x_2}{R} \partial_R n^{\rm g} = A\,\sin\theta\, /\,{n^{\rm g}}\,,
\end{array}
\label{partialderivatives}
\end{equation}
with
\begin{equation}
A=6\,\Bigl(1-2 \frac{f_R}{R}\Bigr)\frac{f_R}{R^2}\,\hat{s}_R^2 -3
\Bigl(1 +  \frac{f_R}{R}\Bigr)\frac{f_R}{R^2}\,\hat{s}_\theta^2 \,,
\label{AR}
\end{equation}
where $f_R$ is given by the  second relation of Eq.~(\ref{tensionhomsphereGreater}).

The Euler method is a first-order method, which means that the local error per step is proportional to the square of the step size, and the global error over the total path is proportional to the step size $\Delta s$. Within this first-order approximation, we may take the virtual refractive index outside the differentiation with respect to $s$ and divide both sides of Eq.~(\ref{raydifeq}) by ${n^{\rm g}}$. To reduce the errors, one may apply a so-called predictor\mbox{-}corrector method. Since the spatial variation of the geodetic line is very small and exhibits only some variation during the passage along the sphere, the extra corrector step is not necessary.
We use a step size of $\Delta s = 0.01 \, R_{\cal S}$. After carrying out step (2a), we update the   values for the virtual refractive index  and its spatial derivatives and  we return to step (1). The recursion is terminated, when the geodetic line has reached the boundary of our window of observation. Step (2a) seems superfluous, but we need the expression in our asymptotic analysis for small $\alpha$.

 The geodetic line is constructed via the following recursive scheme:
\begin{equation}
\begin{array}{lrcl}
{\rm Step}\, (1): &x_2 \!\!&\! :\,=\!&\!\!   x_2 + \sin\alpha \; \Delta s   \,,\\[2mm]
&x_1 \!\!&\! :\,=\!&\!\!   x_1 + \cos\alpha \; \Delta s     \,,  \\ [4mm]
{\rm Step}\, (2): &\hspace*{5mm} \sin\alpha \!\!&\! :\,=\!&\!\!   \sin\alpha  + (\partial_2{n^{\rm g}}/{n^{\rm g}})\, \Delta s \,, \\[2mm]
& \cos\alpha \!\!&\! :\,=\!&\!\! \cos\alpha  + (\partial_1{n^{\rm g}} / {n^{\rm g}}) \, \Delta s     \,,\\[4mm]
{\rm Step}\, {\rm (2a)}: \hspace*{-1cm}& \alpha \!\!&\! :\,=\!&\!\! \arctan\!\left( \displaystyle\frac{ \sin\alpha  + (\partial_2{n^{\rm g}}/{n^{\rm g}})\Delta s}
 {\displaystyle\cos\alpha  + (\partial_1{n^{\rm g}}/ {n^{\rm g}})\, \Delta s )}\right) . \\[5mm]
\end{array}
\label{Eulermethod}
\end{equation}

\begin{figure}
 \centering
\includegraphics[width=0.50\textwidth,viewport = 118 325 486 550, clip = true] {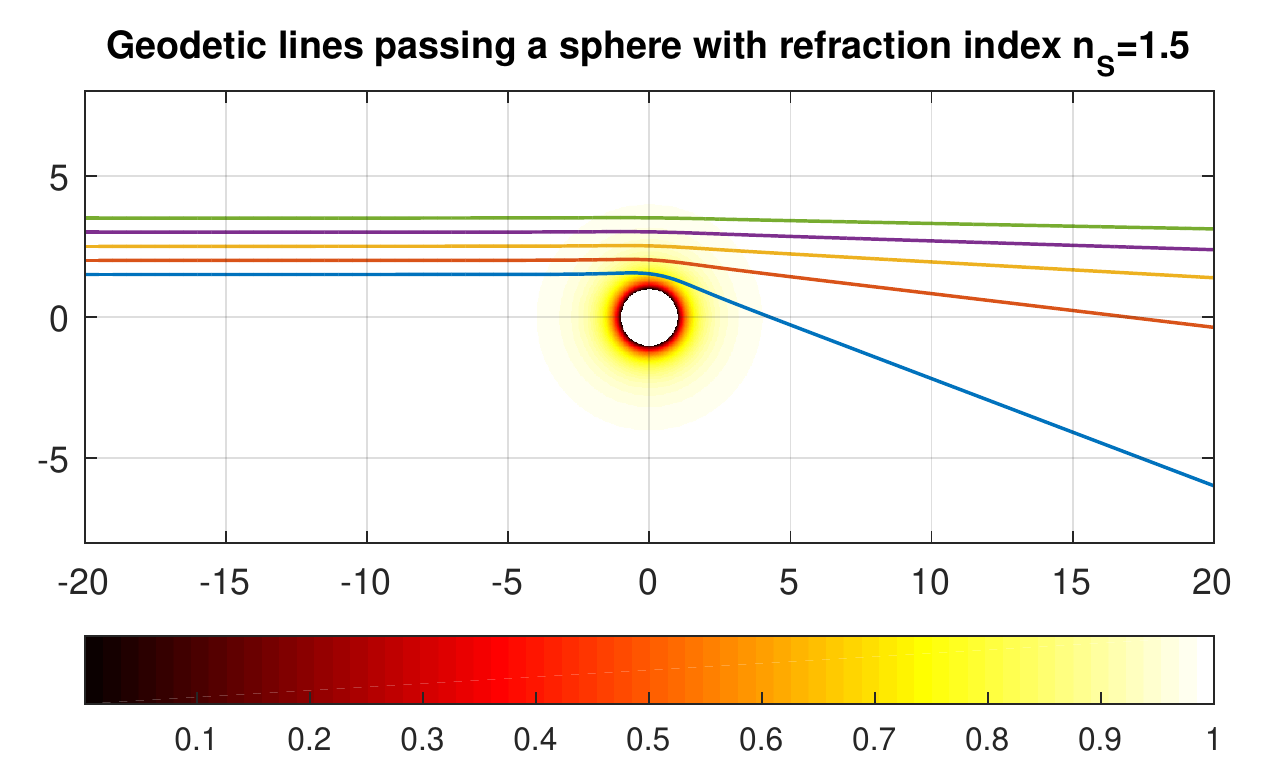}\\[-6mm]
\includegraphics[width=0.50\textwidth,viewport = 118 320 486 550, clip = true] {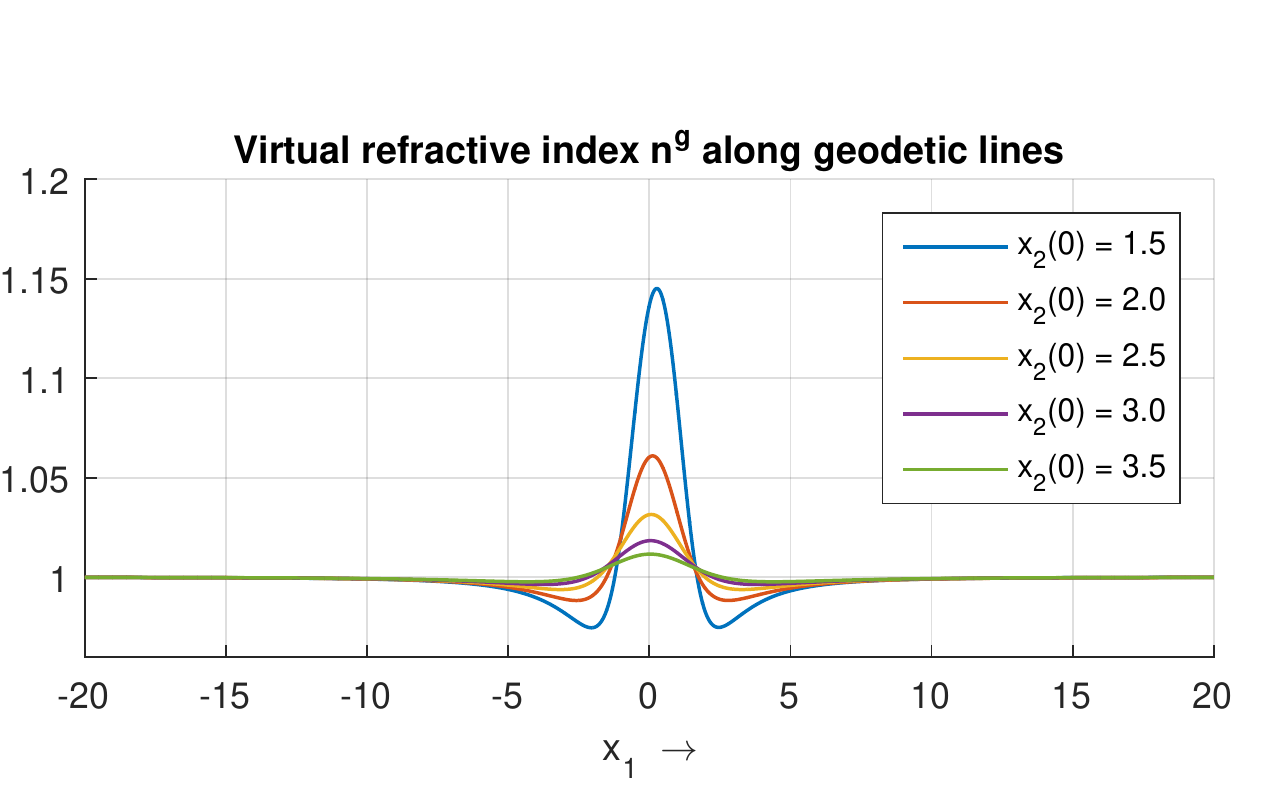}\\[-8mm]
\includegraphics[width=0.50\textwidth,viewport = 118 322 486 550, clip = true] {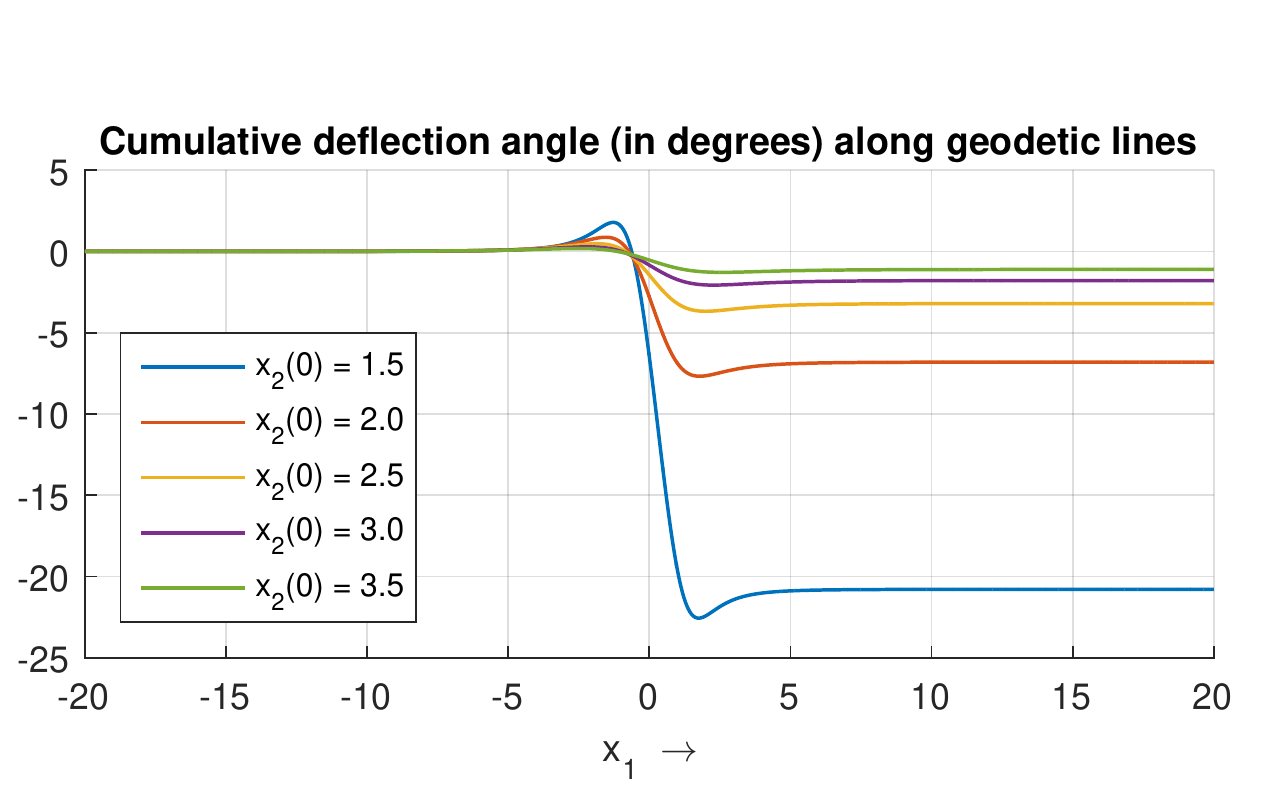}\\[-2mm]
\caption{{\em Top figure}: Geodetic lines in presence of a homogeneous sphere of radius of 1, starting  in the $(x_1,x_2)$-plane at  $x_1(0)= -20$, for various values of $x_2(0)$.  The refractive index $n_S$ of the sphere is 1.5.  All coordinates are normalized with respect to the radius  $R_{\cal S}$.
To indicate the region where the  virtual refractive index is effective, we have  included an image of the quantity $1-(R_{\cal S}/R)^2$. The colorbar indicates these values.
{\em Middle figure}: The virtual refractive index ${n^{\rm g}}$ as function of $x_1$ along the geodetic lines  plotted in the top figure.
{\em Bottom Figure}: The cumulative deflection angles in degrees as function of $x_1$ along the geodetic lines plotted in the top figure.}
\label{Fig:Bending}
\end{figure}

\begin{figure}
 \centering
\includegraphics[width=0.50\textwidth,viewport = 118 325 486 550, clip = true] {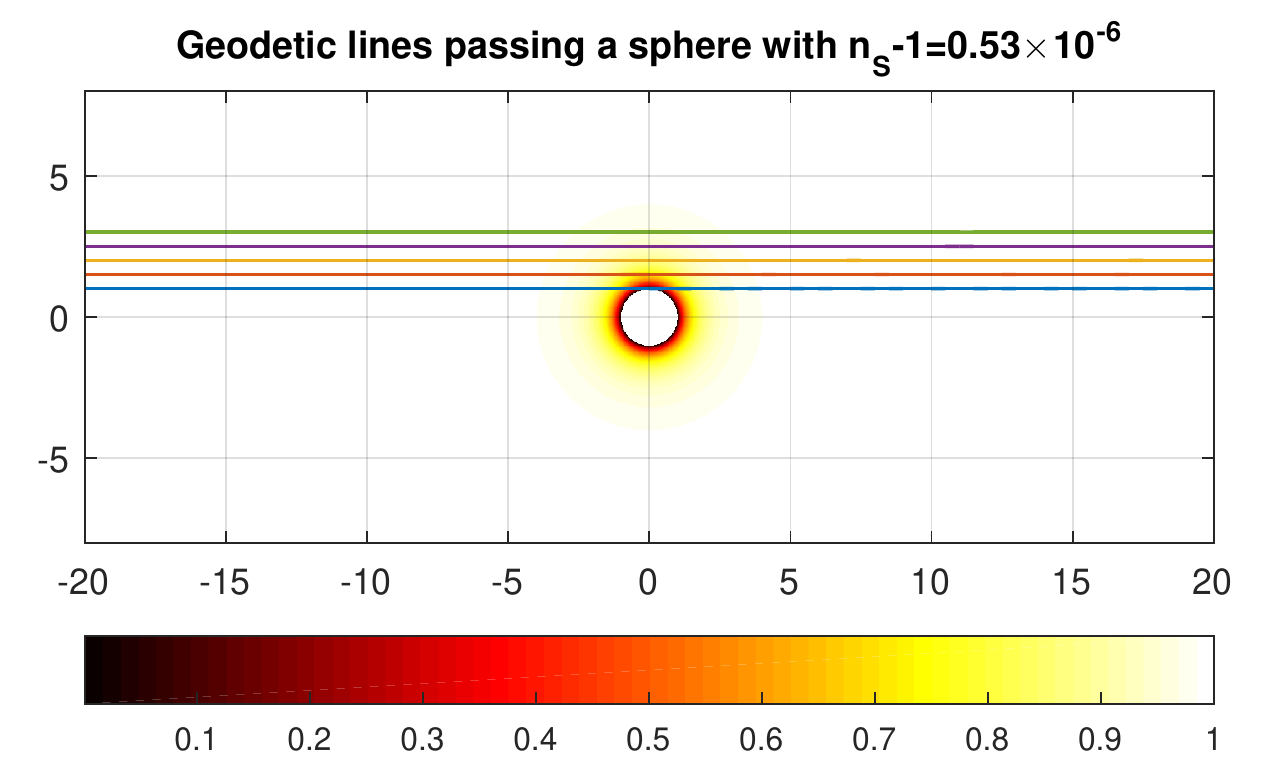}\\[-6mm]
\includegraphics[width=0.50\textwidth,viewport = 118 320 486 550, clip = true] {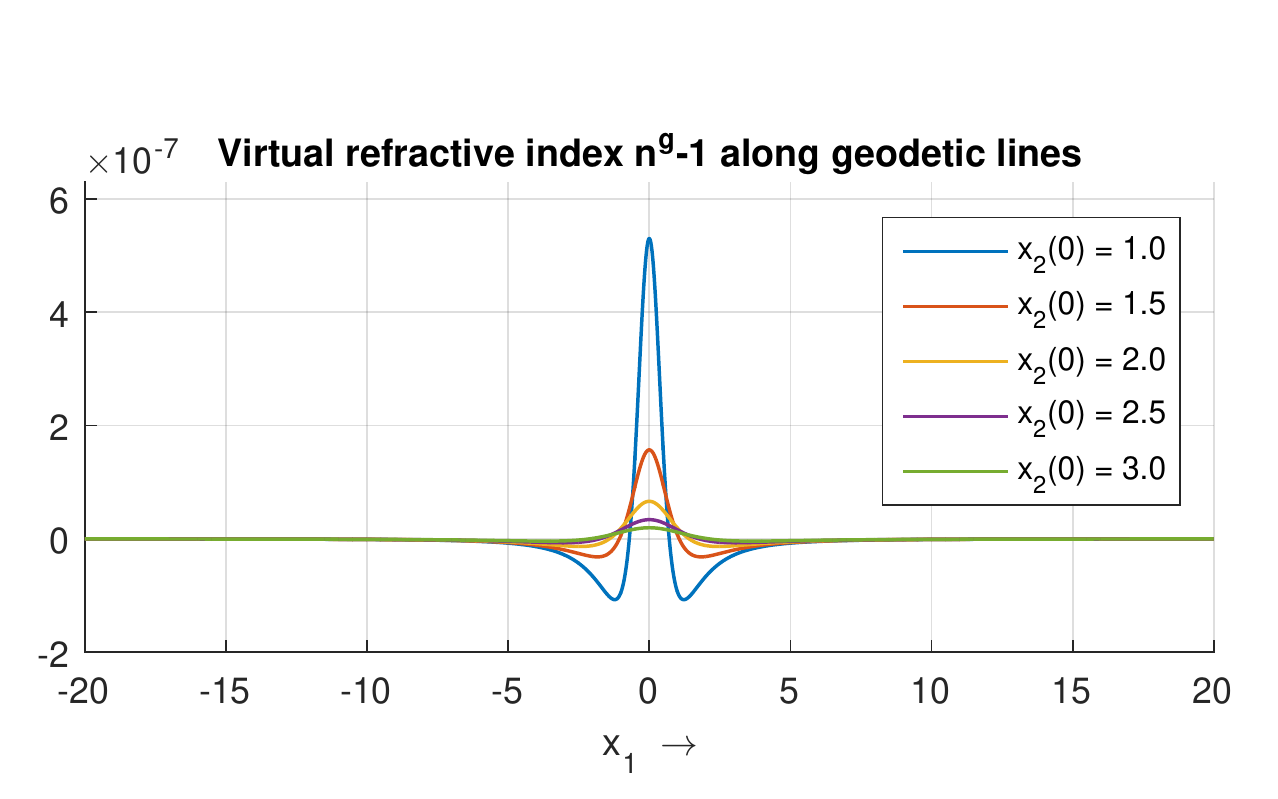}\\[-8mm]
\includegraphics[width=0.50\textwidth,viewport = 118 322 486 550, clip = true] {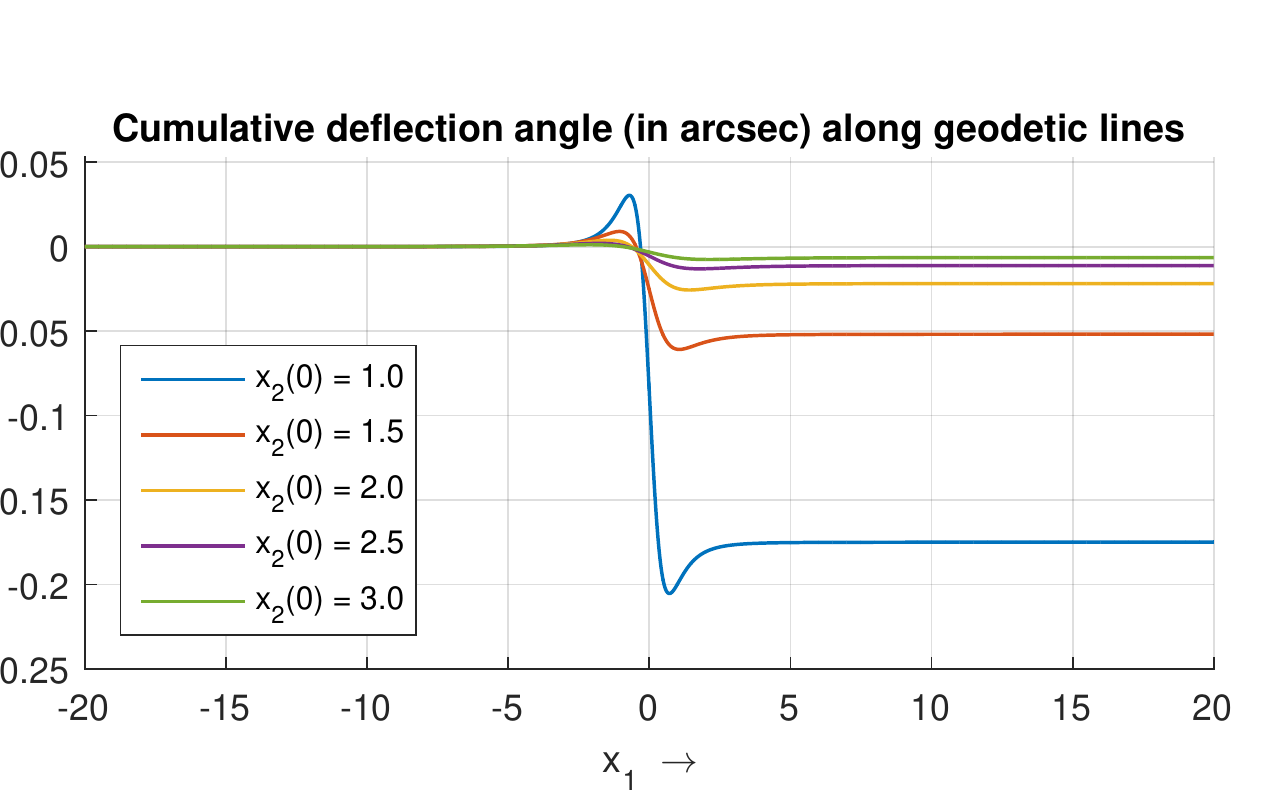}\\[-2mm]
\caption{Same as Fig.~\ref{Fig:Bending}, but now for a refractive index $n_S-1 = 0.53 \times 10^{-6}$ and a different set of $x_2(0)$.  In the middle figure, the vertical axis represents the values of the virtual refraction index ${n^{\rm g}}-1$. In the bottom figure, the angles are given in arcseconds.}
\label{Fig:Small Bending}
\end{figure}

In FIG.~\ref{Fig:Bending}, we show some numerical results of  geodetic lines constructed for $n_{\cal S}=1.5$. In its top figure, the phenomenon of bending of the geodetic lines located  outside the sphere
is clearly visible. We also applied  the construction of these geodetic lines with a
 predictor\mbox{-}corrector method and we did not observe visible differences.
To gain some insight on the influence of the virtual refractive index  on the course of the geodetic path, we picture in the middle figure its value along the path. Approaching the sphere, the virtual velocity $v = c_0/{n^{\rm g}}$ increases and the geodetic line bends away from the sphere. Subsequently, the virtual velocity decreases in the radial direction and the geodetic line bends towards the sphere.
The curvature of the wave fronts in the bottom figure of Fig.~\ref{Fig:Bending} agrees with this phenomenon.
We observe that the presence of the sphere is noticeable in the horizontal range of $(-5R_{\cal S},5R_{\cal S})$. Outside this range the virtual refractive index tends to 1 and the geodetic lines become straight. In the bottom figure, we present the cumulative deflection angles for the different geodetic paths.

In FIG.~\ref{Fig:Small Bending},
we mimic the bending of light by the Sun. The mean refractive index of the Sun is  close to one. For grazing incidence, its value is chosen such that a total deflection angle of 1.75 arcsec is obtained. This is equivalent by taking  $n_{\cal S} = 1 + 0.53\times 10^{-6}$.
A comparison with FIG.~\ref{Fig:Bending} shows that in the top figure of FIG.~\ref{Fig:Small Bending} the deflection of the geodetic lines is hardly visible. The same applies for the shift of the maximum values of the virtual refractive index in the middle figure. The curves are now almost symmetric with respect to $x_1$. In the bottom figure of FIG.~\ref{Fig:Small Bending}, we present the cumulative deflection angle in arcsec. Comparing this picture with the bottom picture of FIG.~\ref{Fig:Bending},  apart of their amplitudes, the global spatial behavior is similar.

\subsection*{Asymptotic analysis for small tension}
\vspace{-1mm}

We take advantage of the very small deflection angles of the geodetic lines outside the sphere. We integrate the differential equation for the geodetic line analytically, after making some appropriate approximations for  refractive indices close to one and small deflection angles.
We start with the expression for $A$ of Eq.~(\ref{AR}). For small values of $n_{\cal S}\!-\!1$, we only retain the terms linear in $f_R$, i.e.,
\begin{equation}
A= (n_{\cal S}\!-\!1)
\left[-3 + 9 \cos^2(\theta\!-\!\alpha)\right]  \frac{R_{\cal S}^3}{R^4}\,.
\label{ARasympt}
\end{equation}
Further, in the region around the sphere, where $ \theta$ has values around $\theta = \frac{1}{2}\pi$, we neglect the influence of $\alpha$. Outside this region, the values of $A$ become negligible.
Next we consider the relation for $\alpha$ of Eq.~(\ref{Eulermethod}).  Within our approximations already made, we take $\sin\alpha \approx \alpha$, $\cos(\theta\!-\!\alpha)  \approx \cos\theta$ and ${n^{\rm g}}\approx 1$. Subsequently,  we expand the  quotient of step (2a) of Eq.~(\ref{Eulermethod}) in terms of small  $\Delta s$ to obtain the cumulative  deflection angle,
\begin{equation}
\alpha : = \alpha + (\partial_2{n^{\rm g}} -\partial_1{n^{\rm g}}) \, \Delta s
\approx  \alpha + \left[ A \sin\theta -  A\cos\theta \right] \Delta s\,,
\label{gamma_asympt}
\end{equation}
where we have used Eq.~(\ref{partialderivatives}).  With $\Delta s \approx \Delta x_1$ and  similar type of approximations made before, the updates for the spatial coordinates become
\begin{equation}
x_2\approx x_2(0) \; \; {\rm and}\; \; x_1 : = x_1 + \Delta x_1\,.
\label{horizontalupdate}
\end{equation}
Since $x_2$ is considered to be constant, we write the radial coordinate as $R = x_2(0)/ \sin\theta$.
The asymptotic expression for $A$ becomes
\begin{equation}
A\approx (n_{\cal S}\!-\!1)
\left[-3 + 9 \cos^2\theta\right]  \frac{R_{\cal S}^3}{x_2^4(0)}\sin^4\theta\,.
\end{equation}
Combining all these approximations, we observe that  Eqs.~(\ref{gamma_asympt}) and (\ref{horizontalupdate}) represent the numerical counterpart to calculate the following integral for the total deflection in the negative $x_2$-direction as:
\begin{equation}
d^{\rm EM} =-\!\int_{-\infty}^\infty  \left[ A \sin\theta -  A \cos\theta \right] {\rm d}x_1 =-2\!\int_0^\infty  \!\! A \sin\theta\, {\rm d}x_1,
\end{equation}
because the first term of the integrand is a symmetric function of $x_1$, while the second  one is asymmetric.
Further, from $x_1 = x_2(0)/ \tan\theta$ follows that $\sin^2\theta\,{\rm d}x_1 = - x_2(0) \, {\rm d}\theta$, and the integral is rewritten as
\begin{equation}
d^{\rm EM} = 2(n_{\cal S}\!-\!1)\left[\frac{R_{\cal S}}{x_2(0)}\right]^3
\int_{0}^{\frac{1}{2}\pi}  \left[3-  9 \cos^2\theta\right] \sin^3\theta\, {\rm d}\theta\,.
\label{DeflIntegral}\end{equation}
The integral can be calculated analytically and is equal to 4/5. The asymptotic formula for the total deflection is finally obtained as
\begin{equation}
d^{\rm EM} = \frac{8}{5} (n_{\cal S}\!-\!1)\frac{1}{(R_0/R_{\cal S})^3}\,,  \quad
\; {\rm for\ small}\; \ n_{\cal S}\!-\!1\,,
\label{gamma_analytic}
\end{equation}
where $R_0 \approx x_2(0)$ is the smallest value of $R$ on the geodetic line. The value of $R_0/R_{\cal S}$
is often denoted as the impact parameter.

\begin{table}[t]
\caption{Ratio $d^{\rm EM}_{\rm num}/d^{\rm EM}_{\rm asymp}$ of numerical values for total deflection angles and their small-tension  approximations}
\label{table:1}
\centering
\begin{tabular}{c cc cc cc }\\[-2mm]
\hline\hline
\\[-2.5mm]
                  &\:\:\:\:\:\:& $\displaystyle n_{\cal S}\!-\!1\!=\!10^{-2}$
                  &\:\:\:\:\:\:& $\displaystyle n_{\cal S}\!-\!1\!=\!10^{-3}$
                  &\:\:\:\:\:\:& $\displaystyle n_{\cal S}\!-\!1\!=\!10^{-4}$ \\
 \\[-2.5mm]
\hline
 \\[-2.5mm]
$R_0/R_{\cal S}$&\:\:\:\:\:\:& 1.02799  &\:\:\:\:\:\:&   1.00278      &\:\:\:\:\:\:&  1.00028             \\
 \\[-2.5mm]
$R_0/R_{\cal S}$&\:\:\:\:\:\:& 1.00831  &\:\:\:\:\:\:&   1.00086      &\:\:\:\:\:\:&  1.00012             \\
\\[-2.5mm]
$R_0/R_{\cal S}$&\:\:\:\:\:\:& 1.00360   &\:\:\:\:\:\:&  1.00046      &\:\:\:\:\:\:&  1.00014             \\
\\[-2.5mm]
$R_0/R_{\cal S}$&\:\:\:\:\:\:& 1.00206   &\:\:\:\:\:\:&  1.00044      &\:\:\:\:\:\:&  1.00028             \\
\\[-2.5mm]
$R_0/R_{\cal S}$&\:\:\:\:\:\:& 1.00158   &\:\:\:\:\:\:&  1.00064      &\:\:\:\:\:\:&  1.00055             \\
 \\[-2.5mm]
\hline                                   
\end{tabular}
\end{table}

\begin{figure*}[t]
 \centering
\includegraphics[width=1\textwidth,viewport = 5 240 600 590, clip = true] {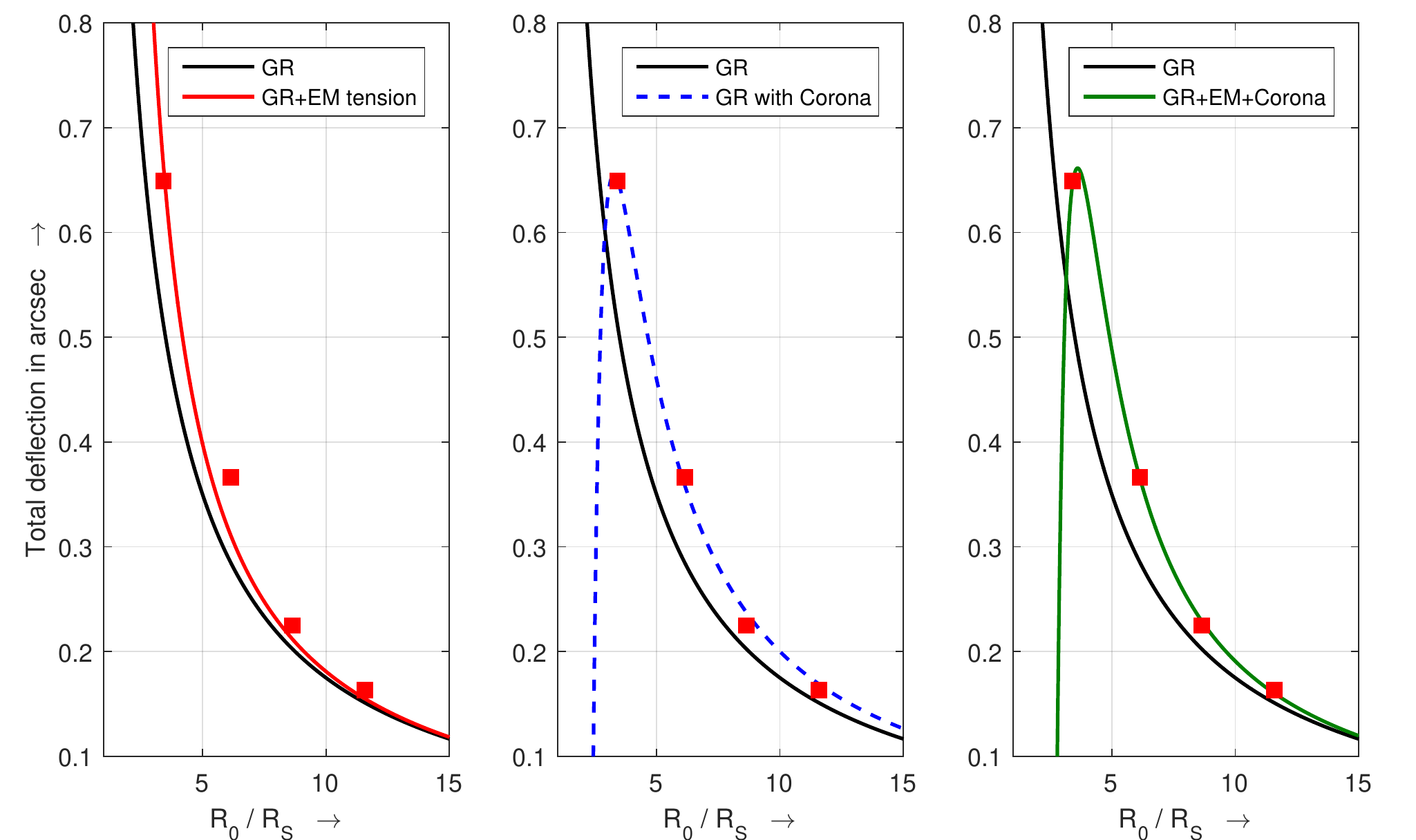}\\
\caption{{\em Left figure}: The GR deflection (black line) and the linear superposition of GR  and  EM deflection  by the tension  of Sun's interior refractive index (red line). The red squares denote the data given by Merat {\em et al} \cite{Merat1974}.
{\em Middle figure}: The linear superposition of  GR deflection and  deflection caused by the Corona (blue dashed line).
{\em Right figure}: The linear superposition of the GR deflection, the EM deflection caused by the tension  of the Sun and the one caused by the Corona (green line).
}
\label{Fig:MeanBending}
\end{figure*}

In TABLE~\ref{table:1}, we present the  values for the ratio of the deflection angle obtained  numerically ($d^{\rm EM}_{\rm num}$) and the one obtained analytically ($d^{\rm EM}_{\rm asymp}$) using the asymptotic approximations. We note that the closer this ratio is to one, the better the asymptotic approach. We observe that the discrepancies in the approximations are of the order of $n_{\cal S}\!-\!1$. For increasing values of $R$ the approximations improve as well; the worst case appears for grazing incidence ($R_0/R_{\cal S}=1$); but we may conclude that, for values of $n_{\cal S}\!-\!1< 10^{-3}$, the approximation is fully justified. Note that for the  Sun case, deflection angles of the order of  1.75 arcsec are arrived at, which correspond to $n_{\cal S}\!-\!1\approx 10^{-6}$. Hence, there is no doubt about the validity of our asymptotic analysis.

From Eq.~(\ref{gamma_analytic}) we conclude that deflection angle is proportional to $(R_0/R_{\cal S})^{-3}$, while general relativity predicts  a dependency of $(R_0/R_{\cal S})^{-1}$, see Fig. 1 of Biswas {\em et al}
\cite{BiswasMani2004}.
For convenience, we denote the electromagnetic contribution, the near-field term of the bending, while the gravitational contribution dominates the
far-field term.

\section{Validation on historical data}

At this point, we return to the work of Merat {\em et al} \cite{Merat1974}. They conclude  on basis of radio deflection observations \cite{Muhleman1970},  that for $ R_0 < 5\, R_{\cal S}$ deviations from the Einstein prediction become statistically significant.
They have collected the whole set of star deflection data into 4 samples. The weighted mean of the distance $R_0/R_{\cal S}$ of each sample  has been given, together with the mean deviation of  light deflections from the GR prediction. In FIG.~\ref{Fig:MeanBending}, the red squares denote the total deflection data values, including the GR predictions.
In the  left picture of FIG.~\ref{Fig:MeanBending}, the GR curve itself is shown as the  black curve. The differences of the data with the GR curve are the four data points given in Table 3 of
Merat {\em et al} \cite{Merat1974}.
For decreasing $R_0/R_{\cal S}$, the discrepancies with the GR prediction increase. The discrepancies between the four data points  and this black curve amounts to {\bf 0.139, 0.081, 0.023 and 0.013}, respectively. The mean square of these residuals with respect to the total deflection error amounts to {\bf 31 \%}.

\subsection{Influence of the EM tension}

To improve the GR reflection model, we assume that the total deviation data is a linear superposition of the GR curve  $(R_0/R_{\cal S})^{-1} $ and our EM curve   $(R_0/R_{\cal S})^{-3} $. We define
\begin{equation}
 d^{\rm EM}= d^{tot} - \frac{1.75 \,{\rm arcsec}}{(R_0/R_{\cal S})}= \frac{B}{(R_0/R_{\cal S})^3}\,.
\label{alphafit}
\end{equation}
To find the unknown factor   $B$, we carry out a least-square fit, which minimizes the deviations and the four data points of \cite{Merat1974}.
Substituting the resulting value of $B$ in Eq.~(\ref{alphafit}), the total deflection function $d^{\rm GR} +d^{\rm EM}$ is presented as the red line in the left picture of FIG.~\ref{Fig:MeanBending}. The discrepancies between the four data points  and this red curve amounts to {\bf -0.0107, 0.0549, 0.0137 and 0.0091}, respectively.
The mean square of these residuals with respect to the total deflection error amounts to {\bf 14 \%}.
These  discrepancies  with respect to the four data points may be explained as a Corona effect outside the domain ${\cal S}$ of the Sun.

\subsection{Influence of the Corona}
\vspace{-1mm}
In the Corona, we only take  into account the local effect of the refractive index of the Corona.
In order to include the plasma effects of the Corona, we start with the refractive index described  as a superposition of  powers of $R_{\cal S}/R$, with constant factors $\eta_p$, viz.
\begin{equation}
n_{\cal C}(R)-1 =  \sum_p\eta_p \Bigl(\frac{R_{\cal S}}{R}\Bigr)^{p}\,,  \quad p>1, \quad R>R_{\cal S}\,.
\end{equation}
The  data under consideration are obtained for $R > 3R_S$ and we employ  the refractive index described in \cite{Muhleman1970}, viz.,
\begin{equation}
n_C(R)-1 = \eta_{p_1} \Bigl(\frac{R_{\cal S}}{R}\Bigr)^{p_1} + \eta_{p_2} \Bigl(\frac{R_{\cal S}}{R}\Bigr)^{p_2}\,,  \quad  \frac{R_{\cal S}}{R} > 3\,,
\end{equation}
where  $p_1 = 6$ and $p_2 =2.33$.
 Using the results of  Appendix B, the electromagnetic deflection is obtained as
\begin{equation}
d^{EM} =\frac{C_{p_1}}{(R_0/R_{\cal S})^{p_1}}
 + \frac{C_{p_2}}{(R_0/R_{\cal S})^{p_2}} \,.
\label{alphafitCorona}
\end{equation}
For the range of $R_0 > 3 R_{\cal S}$ we determine  the coefficients  $C_{p_1}$ and $C_{p_2}$  by a least-square fitting of Eq.~(\ref{alphafitCorona}) to the four data points given by Merat {\em et al} \cite{Merat1974}.

In the middle picture of FIG.~\ref{Fig:MeanBending}, the deflection by the local coronal medium is presented as the blue dashed line. The discrepancies between the four data points and this blue dashed curve are {\bf -0.0002, 0.0071, -0.0119  and -0.0048}, respectively. The mean square of these residuals with respect to the total deflection error amounts to {\bf 5 \%}.

\subsection{Influence of the EM tension and the Corona}
In the integral expression of $f_R$, see Eq.~(\ref{sphere2}), for $R>R_{\cal S}$, we subtract in the integrand $n(r)\!-\!1$ the coronal contribution, $n_{\cal C}(r)\!-\!1$, so that the integral is restricted to
the range of $0<r<R^S$. Then, the EM deflection is given by Eq.~(\ref{alphafit}). Following the pure gravity light bending theory of Maccone \cite{Maccone2009}, we also denote this as the {\em naked-Sun} situation. For small deflections, we take a linear superposition of the {\em naked-Sun} part and the {\em mantle} part (the Corona). We conclude that the total electromagnetic deflection may be written as
\begin{equation}
d^{EM} =\frac{B}{(R_0/R_{\cal S})^3} + \frac{C_{p_1}}{(R_0/R_{\cal S})^{p_1}}
 + \frac{C_{p_2}}{(R_0/R_{\cal S})^{p_2}} \,.
\label{alphaNakedSun}
\end{equation}
In a least-square fitting procedure to the data, we observed that the system matrix is heavily ill-posed and impossible to invert numerically. A stable result is obtained by preconditioning.
We rewrite Eq.~(\ref{alphaNakedSun}) as
\begin{equation}
d^{EM} =\frac{B}{(R_0/R_{\cal S})^3}\left[1 + \frac{C_1}{(R_0/R_{\cal S})^{p_1-3}}
 + \frac{C_2}{(R_0/R_{\cal S})^{p_2-3}}\right] ,
\label{alphafitCoronalmod}
\end{equation}
where $C_1 = C_{p_1} /B$ and $C_2 = C_{p_2}/B$. This nonlinear equation is solved with an iterative Gauss-Newton method. As starting values we take zero values for $C_1$ and $C_2$ and determine $B$ by a direct least-square minimization.
 After carrying out a few Gauss-Newton iterations a stable result is obtained. The resulting deflection is plotted as the green line in the right picture of Fig.~\ref{Fig:MeanBending}.
 The discrepancies
between the four data points and this red curve amounts to
 {\bf -0.0000, 0.0008. -0.0036 and  0.0040}, respectively.
  The mean square of these residuals with respect to the total deflection error amounts to {\bf 2 \%}.
  In comparison with the deviation as sum of the GR and Corona constituents, the present global error
 has been halved by more than a factor of two.

\begin{table}[t]
\caption{Parameters $B$, $C_{p_1}$ and $C_{p_2}$ (in arcsec) substituted into Eq.~(\ref{alphafitCoronalmod}) to plot the three curves of Fig.~\ref{Fig:MeanBendingCorona}.
For convenience we also present the ratio of $C_{p_1}/C_{p_2}$.  }
\label{table:2}
\centering
\begin{tabular}{ccccccccccc}\\[-2mm]
  \hline\hline\\[-2mm]
   && naked Sun  && mantle  && naked Sun $\!+\!$ mantle  \\
   && (red curve)  && (blue dashed)  && (green curve)  \\[2mm]
  \hline\\[-2mm]
  $B$       && 6.04 && 0      &&   36.8 \\[2mm]
  $C_{p_1}$     && 0    && -274    && $-3.17\times 10^4$\\[2mm]
  $C_{p_2}$     && 0    && 5      &&   $-1.59\times 10^2$ \\[2mm]
  $C_{p_1}/C_{p_2}$ && - && -50       && 199 \\[2mm]
  \hline\\
\end{tabular}
\end{table}

\begin{figure}[t]
 \centering
\includegraphics[width=0.48\textwidth,viewport = 115 240 460 590, clip = true] {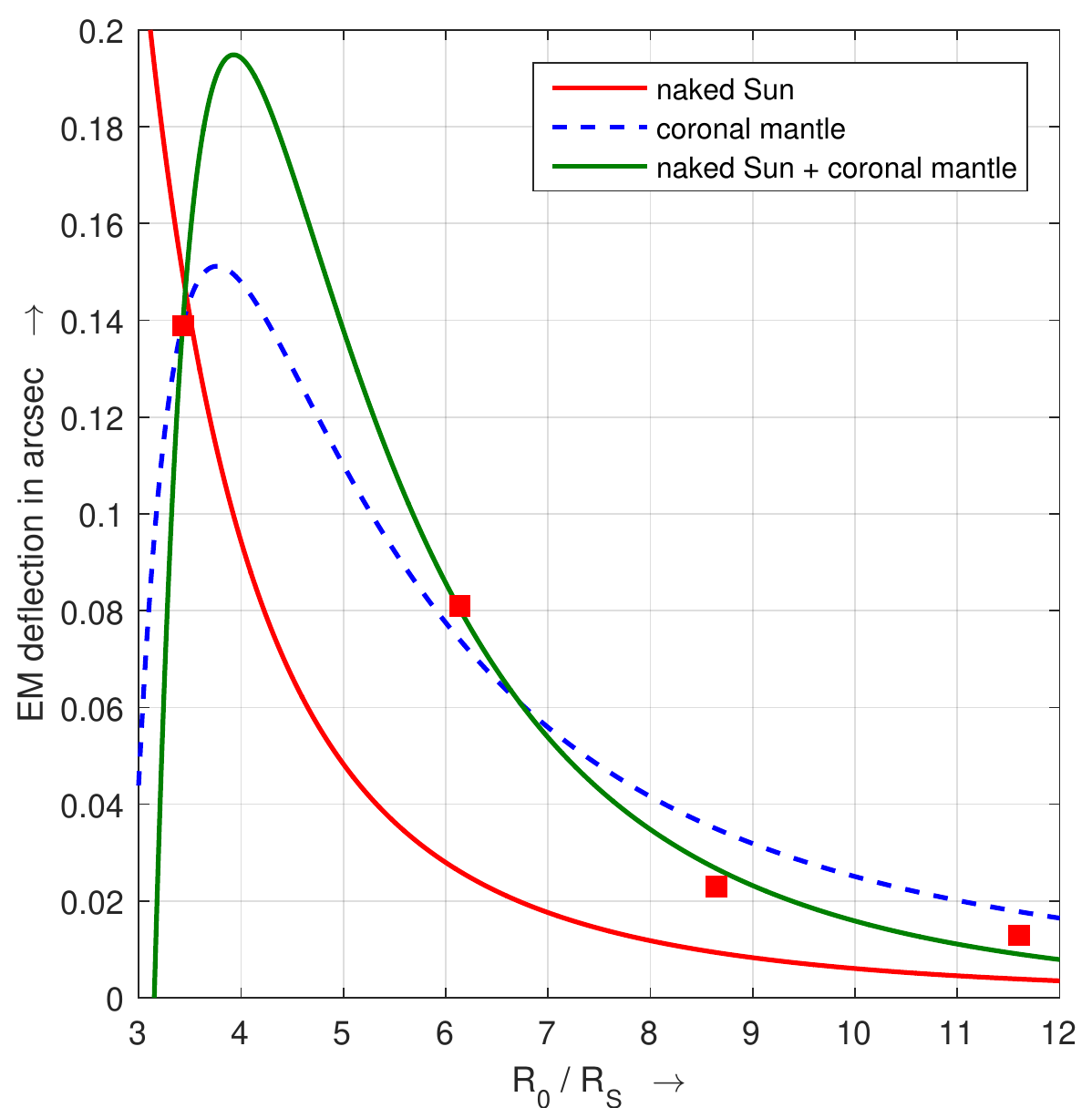}
\caption{Details of the additional EM deflection in a vacuum medium as function of  $R_0/R_{\cal S} $. The red line is the deflection result due to the tension of Sun's interior, without the Corona ({\em naked Sun}). The blue dashed line is the deflection result due to propagation of the wave through the Corona only ({\em coronal mantle}).
The green line denotes the deflection result due to the tension of Sun's interior and the bending of light through a coronal medium ({\em naked Sun + coronal mantle}).
 The red squares denote the data given by Merat {\em et al} \cite{Merat1974}.}
\label{Fig:MeanBendingCorona}
\end{figure}

To judge the value of the different results, in Table~\ref{table:2} we present the values of the three parameters $B$, $C_{p_1}$ and $C_{p_2}$ obtained from our three  fitting procedures.
Also the differences between the results of Fig.~\ref{Fig:MeanBending} becomes more visible when we only present the electromagnetic parts of the deflection $d^{\rm EM}$, see Fig.~\ref{Fig:MeanBendingCorona}.
Although the blue dashed and green curve have a similar shape, their parameters $C_{p_1}$ and $C_{p_2}$
are completely different. Note that the ratio  $C_{p_1}/C_{p_2} = 199$, obtained in the fourth column,  is close to a similar ratio of 228/1.1 = 207, determined empirically by Turyshev and Andersson \cite{TuryshevAndersson2003}. When we neglect the refractional  tension ($B =0$), we observe in the third column a negative  ratio $C_{p_1}/C_{p_2}$, which seems in contradiction to other historical data,  e.g. \cite{Muhleman1970}.
Under condition that we keep the GR prediction unchanged, we claim that the near-field correction  due to the tension of the Sun's interior refractive index is
a prerequisite to obtain an accurate model in solar gravitational lensing \cite{Eshleman1979,Maccone2009}.

\subsection{Frequency dependent bending}
\vspace{-1mm}

Consistent with the frequency dependent refractive index $n_{\cal C}=n_{\cal C}(\omega)$  of the coronal medium
also the  interior refraction index $n_{\cal S}=n_{\cal S}(\omega)$ of the Sun is frequency dependent.
The additional EM deflection
$d^{\rm EM}$ is linearly related to these frequency-dependent refractive indices.
As far as the frequency-dependent  refractive index of Sun's interior is concerned,
the outer layer can be represented by  a refractive index that differs from the other inner layers. This will change its mean value $n_{\cal S}$.

\section{Conclusions}

In this paper we demonstrated that apart from a gravitational type  and a coronal type of bending along the Sun, Maxwell's equations predict an additional type of bending by the existence of a refractional tension. The latter is caused by the presence of a non-zero refractive
 index of the Sun's interior medium. This electromagnetic addition to the  GR tension has been verified on  data from historical astrophysical measurements without changing the GR tension. It has been shown that the additional EM tension is an essential ingredient of the prediction of the interstellar wave propagation paths.
 The influence of the refractional tension becomes more significant for observations closer to the Sun.

The electromagnetic deflection (including the coronal one) is frequency-dependent and dominant in the near- and mid-field, while the GR contribution is frequency independent and it dominates the far-field.  Future research is extremely important for gravity lensing and interstellar communication experiments, where accurate electromagnetic predictions of possible interstellar pathways are sought.

We conclude our paper by mentioning that a scaled experiment is possible by using a voluminous object with a noticeable refractive index. This will potentially verify the electromagnetic deflection outside the object, since the gravitational and coronal components can be neglected in this case.

\begin{acknowledgments}
\vspace{-1mm}
The authors would like to thank Dr. Joost van der Neut for critical review and stimulating discussions.
\end{acknowledgments}

\bibliography{BendingLight}

\begin{thebibliography}{15}%
\makeatletter
\providecommand \@ifxundefined [1]{%
 \@ifx{#1\undefined}
}%
\providecommand \@ifnum [1]{%
 \ifnum #1\expandafter \@firstoftwo
 \else \expandafter \@secondoftwo
 \fi
}%
\providecommand \@ifx [1]{%
 \ifx #1\expandafter \@firstoftwo
 \else \expandafter \@secondoftwo
 \fi
}%
\providecommand \natexlab [1]{#1}%
\providecommand \enquote  [1]{``#1''}%
\providecommand \bibnamefont  [1]{#1}%
\providecommand \bibfnamefont [1]{#1}%
\providecommand \citenamefont [1]{#1}%
\providecommand \href@noop [0]{\@secondoftwo}%
\providecommand \href [0]{\begingroup \@sanitize@url \@href}%
\providecommand \@href[1]{\@@startlink{#1}\@@href}%
\providecommand \@@href[1]{\endgroup#1\@@endlink}%
\providecommand \@sanitize@url [0]{\catcode `\\12\catcode `\$12\catcode
  `\&12\catcode `\#12\catcode `\^12\catcode `\_12\catcode `\%12\relax}%
\providecommand \@@startlink[1]{}%
\providecommand \@@endlink[0]{}%
\providecommand \url  [0]{\begingroup\@sanitize@url \@url }%
\providecommand \@url [1]{\endgroup\@href {#1}{\urlprefix }}%
\providecommand \urlprefix  [0]{URL }%
\providecommand \Eprint [0]{\href }%
\providecommand \doibase [0]{http://dx.doi.org/}%
\providecommand \selectlanguage [0]{\@gobble}%
\providecommand \bibinfo  [0]{\@secondoftwo}%
\providecommand \bibfield  [0]{\@secondoftwo}%
\providecommand \translation [1]{[#1]}%
\providecommand \BibitemOpen [0]{}%
\providecommand \bibitemStop [0]{}%
\providecommand \bibitemNoStop [0]{.\EOS\space}%
\providecommand \EOS [0]{\spacefactor3000\relax}%
\providecommand \BibitemShut  [1]{\csname bibitem#1\endcsname}%
\let\auto@bib@innerbib\@empty
\bibitem [{\citenamefont {Einstein}(1911)}]{Einstein1911}%
  \BibitemOpen
  \bibfield  {author} {\bibinfo {author} {\bibfnamefont {A.}~\bibnamefont
  {Einstein}},\ }\href@noop {} {\bibfield  {journal} {\bibinfo  {journal}
  {Ann.\ d.\ Phys.}\ }\textbf {\bibinfo {volume} {35}},\ \bibinfo {pages} {898}
  (\bibinfo {year} {1911})}\BibitemShut {NoStop}%
\bibitem [{\citenamefont {Dyson}\ \emph {et~al.}(1920)\citenamefont {Dyson},
  \citenamefont {Eddington},\ and\ \citenamefont {Davidson}}]{Dyson1920}%
  \BibitemOpen
  \bibfield  {author} {\bibinfo {author} {\bibfnamefont {F.~W.}\ \bibnamefont
  {Dyson}}, \bibinfo {author} {\bibfnamefont {A.~S.}\ \bibnamefont
  {Eddington}}, \ and\ \bibinfo {author} {\bibfnamefont {C.}~\bibnamefont
  {Davidson}},\ }\href@noop {} {\bibfield  {journal} {\bibinfo  {journal}
  {Phil. Trans. R. Soc. A}\ }\textbf {\bibinfo {volume} {220}},\ \bibinfo
  {pages} {291} (\bibinfo {year} {1920})}\BibitemShut {NoStop}%
\bibitem [{\citenamefont {Einstein}(1916)}]{Einstein1916}%
  \BibitemOpen
  \bibfield  {author} {\bibinfo {author} {\bibfnamefont {A.}~\bibnamefont
  {Einstein}},\ }\href@noop {} {\bibfield  {journal} {\bibinfo  {journal}
  {Ann.\ d.\ Phys.}\ }\textbf {\bibinfo {volume} {49}},\ \bibinfo {pages} {769}
  (\bibinfo {year} {1916})}\BibitemShut {NoStop}%
\bibitem [{\citenamefont {Will}(2015)}]{Will2015}%
  \BibitemOpen
  \bibfield  {author} {\bibinfo {author} {\bibfnamefont {C.~M.}\ \bibnamefont
  {Will}},\ }\href@noop {} {\bibfield  {journal} {\bibinfo  {journal} {Class.
  Quantum Grav.}\ }\textbf {\bibinfo {volume} {32}},\ \bibinfo {pages} {124001}
  (\bibinfo {year} {2015})}\BibitemShut {NoStop}%
\bibitem [{\citenamefont {Woodward}\ and\ \citenamefont
  {Yourgrau}(1970{\natexlab{a}})}]{WoodwardYourgrau1970a}%
  \BibitemOpen
  \bibfield  {author} {\bibinfo {author} {\bibfnamefont {J.}~\bibnamefont
  {Woodward}}\ and\ \bibinfo {author} {\bibfnamefont {W.}~\bibnamefont
  {Yourgrau}},\ }\href@noop {} {\bibfield  {journal} {\bibinfo  {journal}
  {Nature}\ }\textbf {\bibinfo {volume} {226}},\ \bibinfo {pages} {619}
  (\bibinfo {year} {1970}{\natexlab{a}})}\BibitemShut {NoStop}%
\bibitem [{\citenamefont {Woodward}\ and\ \citenamefont
  {Yourgrau}(1970{\natexlab{b}})}]{WoodwardYourgrau1970b}%
  \BibitemOpen
  \bibfield  {author} {\bibinfo {author} {\bibfnamefont {J.}~\bibnamefont
  {Woodward}}\ and\ \bibinfo {author} {\bibfnamefont {W.}~\bibnamefont
  {Yourgrau}},\ }\href@noop {} {\bibfield  {journal} {\bibinfo  {journal} {Ann.
  Phys.}\ }\textbf {\bibinfo {volume} {25}},\ \bibinfo {pages} {334} (\bibinfo
  {year} {1970}{\natexlab{b}})}\BibitemShut {NoStop}%
\bibitem [{\citenamefont {Treder}(1971)}]{Treder1971}%
  \BibitemOpen
  \bibfield  {author} {\bibinfo {author} {\bibfnamefont {H.~J.}\ \bibnamefont
  {Treder}},\ }\href@noop {} {\bibfield  {journal} {\bibinfo  {journal} {Ann.
  Phys.}\ }\textbf {\bibinfo {volume} {27}},\ \bibinfo {pages} {177} (\bibinfo
  {year} {1971})}\BibitemShut {NoStop}%
\bibitem [{\citenamefont {Merat}\ \emph {et~al.}(1974)\citenamefont {Merat},
  \citenamefont {Pecker}, \citenamefont {Vigier},\ and\ \citenamefont
  {Yourgrau}}]{Merat1974}%
  \BibitemOpen
  \bibfield  {author} {\bibinfo {author} {\bibfnamefont {P.}~\bibnamefont
  {Merat}}, \bibinfo {author} {\bibfnamefont {J.~C.}\ \bibnamefont {Pecker}},
  \bibinfo {author} {\bibfnamefont {J.~P.}\ \bibnamefont {Vigier}}, \ and\
  \bibinfo {author} {\bibfnamefont {W.}~\bibnamefont {Yourgrau}},\ }\href@noop
  {} {\bibfield  {journal} {\bibinfo  {journal} {Astron. \& Astrophys.}\
  }\textbf {\bibinfo {volume} {32}},\ \bibinfo {pages} {471} (\bibinfo {year}
  {1974})}\BibitemShut {NoStop}%
\bibitem [{\citenamefont {Feynman}()}]{Feynman1964}%
  \BibitemOpen
  \bibfield  {author} {\bibinfo {author} {\bibfnamefont {R.~P.}\ \bibnamefont
  {Feynman}},\ }\href@noop {} {\emph {\bibinfo {title} {The Feynman Lectures on
  Physics (\rm{Addison-Wesley} 1964), Volume 1, Chapter 26-5}}},\ \bibinfo
  {note} {see http://www.feynmanlectures.caltech.edu}\BibitemShut {NoStop}%
\bibitem [{\citenamefont {Born}\ and\ \citenamefont
  {Wolf}(1959)}]{BornWolf1959}%
  \BibitemOpen
  \bibfield  {author} {\bibinfo {author} {\bibfnamefont {M.~A.}\ \bibnamefont
  {Born}}\ and\ \bibinfo {author} {\bibfnamefont {E.}~\bibnamefont {Wolf}},\
  }\href@noop {} {\emph {\bibinfo {title} {Principles of Optics}}}\ (\bibinfo
  {publisher} {Pergamon Press},\ \bibinfo {year} {1959})\ pp.\ \bibinfo {pages}
  {121--122}\BibitemShut {NoStop}%
\bibitem [{\citenamefont {Biswas}\ and\ \citenamefont
  {Mani}(2004)}]{BiswasMani2004}%
  \BibitemOpen
  \bibfield  {author} {\bibinfo {author} {\bibfnamefont {A.}~\bibnamefont
  {Biswas}}\ and\ \bibinfo {author} {\bibfnamefont {K.}~\bibnamefont {Mani}},\
  }\href@noop {} {\bibfield  {journal} {\bibinfo  {journal} {Cent. Eur. J.
  Phys.}\ }\textbf {\bibinfo {volume} {25}},\ \bibinfo {pages} {1} (\bibinfo
  {year} {2004})},\ \bibinfo {note} {docID = 10.2478/BF02475569}\BibitemShut
  {NoStop}%
\bibitem [{\citenamefont {Muhleman}\ \emph {et~al.}(1970)\citenamefont
  {Muhleman}, \citenamefont {Ekers},\ and\ \citenamefont
  {Fomalont}}]{Muhleman1970}%
  \BibitemOpen
  \bibfield  {author} {\bibinfo {author} {\bibfnamefont {D.~O.}\ \bibnamefont
  {Muhleman}}, \bibinfo {author} {\bibfnamefont {R.~D.}\ \bibnamefont {Ekers}},
  \ and\ \bibinfo {author} {\bibfnamefont {E.~B.}\ \bibnamefont {Fomalont}},\
  }\href@noop {} {\bibfield  {journal} {\bibinfo  {journal} {Phys. Rev.
  Letters}\ }\textbf {\bibinfo {volume} {24}},\ \bibinfo {pages} {1377}
  (\bibinfo {year} {1970})}\BibitemShut {NoStop}%
\bibitem [{\citenamefont {Maccone}(2009)}]{Maccone2009}%
  \BibitemOpen
  \bibfield  {author} {\bibinfo {author} {\bibfnamefont {C.}~\bibnamefont
  {Maccone}},\ }\href@noop {} {\emph {\bibinfo {title} {Deep Space Flight and
  Communications, Exploiting the Sun as a Gravitational Lens}}}\ (\bibinfo
  {publisher} {Springer},\ \bibinfo {year} {2009})\ pp.\ \bibinfo {pages}
  {113--134}\BibitemShut {NoStop}%
\bibitem [{\citenamefont {Turyshev}\ and\ \citenamefont
  {Andersson}(2003)}]{TuryshevAndersson2003}%
  \BibitemOpen
  \bibfield  {author} {\bibinfo {author} {\bibfnamefont {S.~G.}\ \bibnamefont
  {Turyshev}}\ and\ \bibinfo {author} {\bibfnamefont {B.~G.}\ \bibnamefont
  {Andersson}},\ }\href@noop {} {\bibfield  {journal} {\bibinfo  {journal}
  {Mon. Not. R. Astron. Soc.}\ }\textbf {\bibinfo {volume} {341}},\ \bibinfo
  {pages} {577} (\bibinfo {year} {2003})}\BibitemShut {NoStop}%
\bibitem [{\citenamefont {Eshleman}(1979)}]{Eshleman1979}%
  \BibitemOpen
  \bibfield  {author} {\bibinfo {author} {\bibfnamefont {V.~R.}\ \bibnamefont
  {Eshleman}},\ }\href@noop {} {\bibfield  {journal} {\bibinfo  {journal}
  {Science}\ }\textbf {\bibinfo {volume} {205}},\ \bibinfo {pages} {1133}
  (\bibinfo {year} {1979})}\BibitemShut {NoStop}%
\end{thebibliography}%

\begin{appendix}
\section{The refractional potential and the tension for a radially inhomogeneous sphere and its derivatives}
 For a radially inhomogeneous sphere, the refractional potential of Eq.~(\ref{PHI}) can be calculated analytically.
 We introduce spherical coordinates for the observation points $\Bx$ as
 \begin{equation}
 x_1 = R \sin\theta\cos\phi\,,\:  x_2 = R \sin\theta\sin\phi\,, \: x_3 = R \cos\theta\,, \:
 \end{equation}
 and   spherical coordinates for the integration points $\Bx'$ as
 \begin{equation}
 x_1^\prime = r \sin\theta^\prime \cos\phi^\prime \,,\:  x_2 = r \sin\theta^\prime \sin\phi^\prime \,, \: x_3 = r \cos\theta^\prime \,.
 \end{equation}
 For convenience we  take the polar axis in the direction of $\Bx$.
 Then, the Cartesian distance  and the volume element become
 \begin{equation}
 \begin{array}{rcl}
 |\Bx-\Bx^\prime| &=& [R^2 +r^2 - 2\,R\,r \cos\theta^\prime]^\frac{1}{2}\,,\\[2mm] {\rm dV} & =& r^2 \sin\theta^\prime \,{\rm d}r \,{\rm d}\theta^\prime\,{\rm d}\phi^\prime\,.
 \end{array}
 \end{equation}
 In the resulting integral we first carry out the integration with respect to $\phi^\prime$; this merely amounts to a multiplication by a factor of $2\pi$, so that Eq.~(\ref{PHI}) transfers into
\begin{eqnarray}
\hspace*{-1cm}
\Phi(R,\theta, \phi) &=& \frac{3}{2}\int_{0}^{\infty}\![n(r)-1]\,{\rm } r^2\,{\rm d}r \nonumber\\
&& \hspace*{0.5cm} \int_{0}^{\pi}\!\! \frac{\sin\theta^\prime}
                                                 {[R^2\!+\!r^2 \!-\! 2Rr \cos\theta^\prime]^\frac{1}{2}} {\rm d}\theta^\prime\,.\nonumber
\end{eqnarray}
Next we carry out the integration with respect to $\theta^\prime$, which is elementary. After this, we have
\begin{equation}
\Phi(R,\theta, \phi) = \frac{3}{2}\int_{0}^{\infty} [n(r)-1]\, r^2
                          \left[\frac{R+r}{Rr} - \frac{|R-r|}{Rr}\right]
\,{\rm d}r\,,
\end{equation}
which shows that $\Phi$ is independent of $\theta$ and $\phi$.   Taking into account the meaning of $|R-r|$, we obtain
\begin{equation}
\Phi(R) = \frac{3}{R} \int_{0}^{R}{\rm }[n(r)-1]\, r^2 \, {\rm d}r
 +3\int_{R}^{\infty}[n(r)-1] \,r
\,{\rm d}r\,.
\label{PHI_R}
\end{equation}
Note that this expression holds for all $R$, if $n(r)= O(r^{-2})$ when $r$ tends to infinity.

The gradient of the potential is directed in the radial direction. Hence $\nabla\Phi = (d \Phi / dR)\,{\bf i}_R $. Applying Leibniz' rule for differentiation of an integral to Eq.~(\ref{PHI_R}) yields
\begin{eqnarray}
\hspace*{-1cm}
\frac{d\Phi} {dR}&=& - \frac{3}{R^2} \int_{0}^{R}[n(r)-1]\, r^2 \, {\rm d}r \nonumber\\
&\:\:\:\:\:\:& \hspace*{1cm}+ \frac{3}{R} [n(R)-1] \,R^2 - 3 [n(R)-1] \,R\,,
\label{Leibniz1}
\end{eqnarray}
which simplifies to
\begin{equation}
\frac{d\Phi} {d R}= - \frac{3}{R^2} \int_{0}^{R}[n(r)-1]\, r^2 \, {\rm d}r\,.
\label{Leibniz2}
\end{equation}
With this result, the tension $\Bf = f_R {\bf i}_R = - ({d\Phi}/ {dR})\,{\bf i}_R$ is obtained as
\begin{equation}
f_R(R)=  \frac{3}{R^2} \int_{0}^{R}[n(r)-1]\, r^2 \, {\rm d}r\,.
\label{APPNfR}
\end{equation}

\section{Deflection due to presence of the Corona}

Let us consider the coronal refractive index for a particular term in which the radial dependence is given a  certain inverse power of $p$, viz.,
\begin{equation}
n_C(R) = 1+ \eta_p \Bigl(\frac{R_{\cal S}}{R}\Bigr)^{p}\,,
\end{equation}
Then, the partial derivatives with respect to $x_1$ and $x_2$ are obtained as
\begin{equation}
\partial_1 n_C = \frac{x_1}{R} \partial_R n_C\,, \quad
\partial_2 n_C = \frac{x_2}{R} \partial_R n_C\,,
\end{equation}
where
\begin{equation}
\partial_R n_C(R) =  - \eta_p \,p \,\frac{R_{\cal S}^{p}}{R^{p+1}}\,.
\end{equation}
Similar as before, for small values of $\eta$,  the cumulative  deflection angle is given by
\begin{eqnarray}
\alpha &:& = \alpha + (\partial_2 n_C -\partial_1 n_C) \, \Delta s\nonumber\\[2mm]
&&=  \alpha - \eta_p\, p \,\frac{R_{\cal S}^{p}}{R^{p+1}}(\sin\theta - \cos\theta ) \Delta s\,,
\end{eqnarray}
and the total deflection caused by the presence of the Corona becomes
\begin{equation}
d^{\rm EM} = -\eta_p\,p \left[\frac{R_{\cal S}}{x_2(0)}\right]^p F(p)\,,
\end{equation}
in which
\begin{eqnarray}
F(p) &=&
\int_{0}^{\pi} \! \! \!\left[\sin^p\!\theta \!- \!\sin^{p}\,\!\!\theta\,\cos\theta\right]\! {\rm d}\theta=
2\!\int_{0}^{\frac{\pi}{2}}\!\! \sin^p\!\theta \, {\rm d}\theta \nonumber\\
&=&
\frac{\sqrt{\pi}\,\Gamma(\frac{1}{2}+\frac{1}{2}p)}{\Gamma(1+\frac{1}{2}p)}\,.
\end{eqnarray}

\end{appendix}

\end{document}